\documentclass[draft,showpacs,eqsecnum,nofootinbib,aps]{revtex4}
\renewcommand{\theequation}{\arabic{equation}}
\def\beq{\begin{equation}}
\def\eeq{\end{equation}}
\def\bea{\begin{eqnarray}}
\def\eea{\end{eqnarray}}

\begin{document}
\title{Gauge symmetry enhancement in Hamiltonian formalism }
\author{Soon-Tae Hong}
\email{soonhong@ewha.ac.kr} \affiliation{Department of Science
Education, Ewha Womans University, Seoul, 120-750 Korea}
\author{Joohan Lee}
\email{joohan@kerr.uos.ac.kr} \affiliation{Department of Physics,
University of Seoul, Seoul 130-743, Korea}
\author{Tae Hoon Lee}
\email{thlee@physics.ssu.ac.kr} \affiliation{Department of
Physics, Soongsil University, Seoul 156-743, Korea}
\author{Phillial Oh}
\email{ploh@newton.skku.ac.kr} \affiliation{Department of Physics,
Sungkyunkwan University, Suwon 440-746, Korea}

\date{\today}%
\begin{abstract}
We study the Hamiltonian structure of the gauge symmetry
enhancement in the enlarged $CP(N)$ model coupled with $U(2)$
Chern-Simons term, which contains a free parameter governing
explicit symmetry breaking and symmetry enhancement. After giving
a general discussion of the geometry of constrained phase space
suitable for the  enhancement, we explicitly perform the Dirac
analysis of our model and compute the Dirac brackets for the
symmetry enhanced and broken cases. We also discuss some related
issues.
\end{abstract}
 \maketitle

\section{Introduction}
\setcounter{equation}{0}
\renewcommand{\theequation}{\arabic{section}.\arabic{equation}}
It is well-known that the nonlinear sigma models exhibit many
interesting physical properties in the large-$N$ limit
\cite{cole}. One of them is the phenomenon of dynamical generation
of gauge boson in $CP(N)$ model ~\cite{golo}, where the auxiliary
$U(1)$ gauge field becomes dynamical through the radiative
corrections \cite{dada}. Recently, some new properties have been
explored in relation with this phenomenon. In particular, in Ref.
\cite{ior} the issue of dynamical generation of gauge boson has
been analyzed in the context of an enlarged $CP(N)$ model in lower
dimensions. In this model, two complex projective spaces with
different coupling constants couple with each other through
interactions which preserve the exchange of the two spaces. In
addition to the two auxiliary $U(1)$ gauge fields (corresponding
to the diagonal $a_\mu$ and $b_\mu$ fields of (\ref{gcom}) below)
which represent each complex projective space, one extra
auxiliary complex gauge field (the off-diagonal $c_\mu$ field of
(\ref{gcom})) is introduced to couple the two spaces in the way
which preserves the exchange symmetry. It turns out that when the
two coupling constants are equal (which corresponds to the case of
$r=1$ of (\ref{rmatrix})), the classical enlarged model becomes
the nonlinear sigma model with the target space of Grassmannian
manifold \cite{gava}. It was shown in Ref. \cite{ior} that the
additional gauge field, $c_\mu$, also becomes dynamical through
radiative corrections. Moreover, in the self-dual limit where the
two running coupling constants become equal, it becomes massless
and combine with the two $U(1)$ gauge fields to yield the $U(2)$
Yang-Mills theory. That is, the gauge symmetry enhancement has
occurred in the self-dual limit. Away from this limit, the complex
gauge field becomes massive and the symmetry remains to be
$U(1)\times U(1)$.

The parameter $r$ could be understood as an explicit gauge
symmetry breaking parameter from $U(2)$ to $U(1)\times U(1)$, with
the mass of the $c_\mu$ field being induced radiatively through
the loop corrections when the symmetry is broken. This could
provide a scheme of generating mass of the gauge bosons.
Therefore, it would be worthwhile to study the enlarged $CP(N)$
model from different aspects. In this paper, we study this model
in the Hamiltonian formulation. We first recall that the gauge
symmetry is realized as the Gauss law type of constraints in the
Hamiltonian method. In the enlarged model of Ref.~\cite{ior}, the
original gauge fields are auxiliary fields which become dynamical
through the quantum corrections. From the Hamiltonian point of
view, these auxiliary fields could be completely eliminated
through the equations of motion from the beginning, and the Gauss
law constraints could be only implicitly realized. However, in
order to see the structure of gauge symmetry more explicitly, we
couple the enlarged $CP(N)$ model with some external gauge fields,
which we choose to be described by the $U(2)$ Chern-Simons term.
Then, we perform the Dirac analysis \cite{dira} of the resulting
theory. The theory has both first and second class constraints,
and it is found that for $r=1$ the Gauss constraints satisfy
$U(2)$ symmetry algebra, whereas for $r\neq 1$ only $U(1)\times
U(1)$ algebra. What happens is that two of the first class
constraints generating the gauge symmetry become second class
constraints away from the self-dual limit, reducing the resulting
gauge symmetry.

However, it turns out that a smooth extrapolation from the
$U(1)\times U(1)$ to $U(2)$ gauge symmetry algebra is not possible
in the Dirac analysis. The reason is that in the Dirac method we
have to compute the inverse of the Dirac matrix which is
constructed with second class constraints only. This inverse
matrix with parameter $r$ becomes singular if we take the limit of
$r \rightarrow 1$, because two of the constraints change from
second class into first class. When this happens, the Dirac matrix
becomes degenerate and the inverse does not exist. From physical
point of view, this singular behaviour could be associated with
the second order phase transition which one encounters in going to
the limit $r=1$ \cite{ior}.

The organization of the paper is as follows. In Sec. 2, we define
the enlarged CP(N) model coupled with Chern-Simons term and
perform the canonical analysis. In Sec. 3, we give a somewhat
general discussion of the geometry of the constrained phase space
suited for gauge symmetry enhancement. In Sec.4, we give an
explicit computation of the Dirac bracket in the case of $r=1$
and $r\neq 1$ separately. Sec. 5 contains conclusion  and
discussions.

\section{The Model}
\setcounter{equation}{0}
\renewcommand{\theequation}{\arabic{section}.\arabic{equation}}

We start from the Lagrangian written in terms of the $N\times2$
matrix $\psi$ such that
\begin{equation}
{\cal L}=\frac{1}{g^2}{\rm tr} \left[(D_\mu \psi)^\dagger (D^\mu
\psi)-\lambda(\psi^\dagger \psi-R)\right] +{\cal L}_{\mbox{cs}},
\label{lag1}
\end{equation}
where the field, $\psi$, is made of two complex $N$-vectors
$\psi_1$ and $\psi_2$ such that
\begin{equation}
\psi=\left[\psi_1, \psi_2\right],~~~
\psi^\dagger=\left[{\psi_1^\dagger \atop \psi_2^\dagger}\right],
\end{equation}
and the hermitian $2\times2$ matrix $\lambda$ is a Lagrange
multiplier. The $2\times2$ matrix $R$ is given by
\begin{equation}
R=\left[\,{r \atop 0}\quad{0 \atop r^{-1}}\right], \label{rmatrix}
\end{equation}
with a real positive $r$. We will also use the notation $R_{ab}=
r_a\delta_{ab}~(a,b, \cdots=1,2)$ with $r_1=r, r_2=r^{-1}$. The
covariant derivative is defined
 as $D_\mu \psi \equiv\partial_\mu \psi
-\psi A_\mu$ with a $2\times2$ anti-hermitian matrix gauge
potential $A_\mu$ associated with the local $U(2)$ gauge
transformations. The components of $A_\mu$ can be explicitly
written as follows;
 \beq A_\mu =-i\left[
\begin{array}{ll}
a_{\mu} & c_{\mu}^{*}\\
c_{\mu} &b_{\mu}\\
\end{array}
\right].
\label{gcom}
\eeq
${\cal L}_{\mbox{cs}}$ is the
non-Abelian Chern-Simons gauge action given by
\begin{equation}
{\cal L}_{\mbox{cs}}=-\frac{\kappa}{2}\epsilon^{\mu\nu\rho} {\rm
tr} \left(\partial_\mu A_\nu A_\rho+\frac{2}{3}A_\mu A_\nu A_\rho
\right).
\end{equation}
The kinetic term of the Lagrangian (\ref{lag1}) is invariant under
the local $U(2)$ transformation, while the matrix $R$ with
$r\neq1$ explicitly breaks the $U(2)$ gauge symmetry down to $U(1)
\times U(1)$. Thus, the symmetry of our model is $[SU(N)]_{\rm
global} \times [U(2)]_{\rm local}$ for $r=1$, while $[SU(N)]_{\rm
global} \times [U(1) \times U(1)]_{\rm local}$ for $r\neq 1$.
Therefore, the parameter $r$ could be regarded as an explicit
symmetry breaking parameter.

Let us perform the canonical analysis using the Dirac
method~\cite{dira}.  We first define the conjugate
momenta of the $\psi_a^\alpha$ field by
$\Pi_a^\alpha=\frac{\partial {\cal L}}{\partial \dot \psi_a^\alpha}$,
which gives
\begin{eqnarray}
\Pi_a^\alpha=\frac{1}{g^2}(\dot\psi_a^{\alpha\dagger}+A_{0ab}\psi_b^{\alpha\dagger}).
\end{eqnarray}
The indices $a,b...$ represent the $U(2)$ indices 1 and 2, while
Latin indices $\alpha, \beta..$ represent the global $SU(N)$
indices of $\psi_1$ and $\psi_2$. We will occasionally omit the
global $SU(N)$ indices, when the context is clear. Likewise, the
conjugate momentum of the $\psi^{\alpha\dagger}_a$ field is given
by
\begin{eqnarray}
\Pi_a^{\alpha\dagger}=\frac{1}{g^2}(\dot \psi_a^\alpha-
\psi_b^\alpha A_{0ba}).
\end{eqnarray}
The momentum for the Lagrangian multiplier field $\lambda_{ab}$ is
constrained to vanish,
\begin{eqnarray}
\Pi^{\lambda}_{ab}=0. \label{lagrange}
\end{eqnarray}
The conjugate momentum $P^\mu_{ab}$ for the gauge field $A_{\mu ab}$
is given by
\begin{eqnarray}
P_{iab}=\kappa\epsilon_{ij}A_{jba},~~ P_{0ab}=0.
\end{eqnarray}
In the above, the indices $i,j,..$ represent the spatial ones with
1 and 2. In the following analysis we will not treat the first
equation as a constraint. Instead $P_{iab}$ is removed from the
beginning and replaced by $\kappa\epsilon_{ij}A_{jba}$
\cite{fadd}. The second equation, together with (\ref{lagrange}),
defines the primary constraint of the theory. The Poisson bracket
is defined by
\begin{eqnarray}
\{\psi_a^\alpha (x),~ \Pi_b^\beta(y)\}
&=&\delta_{ab}\delta^{\alpha\beta}\delta(x-y),\nonumber\\
\{\lambda_{ab}(x),\Pi_{cd}^{\lambda}(y)\}
&=&\delta_{ac}\delta_{bd}\delta(x-y),\nonumber\\
\{A_{0ab}(x),P_{0cd}(y)\}&=&\delta_{ac}\delta_{bd}\delta(x-y)\nonumber\\
\{A_{iab}(x), A_{jcd}(y)\}&=&-\frac{1}{\kappa}\epsilon_{ij}
\delta_{ad}\delta_{bc}\delta(x-y).
\label{canonical}
\end{eqnarray}

After a straightforward Dirac analysis, we find that the system is
described by the canonical Hamiltonian given by \beq {\cal
H}_0=g^2 \Pi_a \Pi_a^\dagger + \frac{1}{g^2}(D_i \psi)_a^\dagger
(D_i \psi)_a
+\frac{1}{g^{2}}\lambda_{ab}(\psi_{b}^{\dagger}\psi_{a}-R_{ba})
+(\Pi_{a}\psi_{b}-\psi_{a}^{\dagger}\Pi_{b}^{\dagger}+\kappa
F_{12ab})A_{0ba}, \eeq
where we denote $F G\equiv F^\alpha
G^\alpha$ and $F_{12 ab}$ is the magnetic field given by
\begin{equation}
F_{12 ab}=\partial_1 A_{2ab}-\partial_2 A_{1ab}+[A_1, A_2]_{ab}.
\end{equation}
Including all secondary constraints, we find that the dynamics is
governed by the following constraints;
\begin{eqnarray}
C^{(0)}_{ab}&=&\Pi^{\lambda}_{ab}\approx 0,\nonumber\\
C^{(1)}_{ab}&=&P_{ab}^0\approx 0,\nonumber\\
C^{(2)}_{ab}&=&\psi_a^\dagger \psi_b- R_{ab}\approx 0,\nonumber\\
C^{(3)}_{ab}&=&\Pi_a \psi_b-\psi_a^{\dagger}\Pi_b^{\dagger}
+\kappa F_{12 ab}\approx 0,\nonumber\\
C^{(4)}_{ab}&=&\Pi_a \psi_b+\psi^\dagger_a \Pi^\dagger_b
-\frac{1}{g^2}[A_0, R]_{ab} \approx 0.
\label{eqq}
\end{eqnarray}
One can check that the time evolution of the above constraints
are closed with a total Hamiltonian ${\cal H}_T={\cal H}_0+
\sum_{u=0}^4\Lambda^{(u)}_{ab}C^{(u)}_{ab}$ using the relations
(\ref{canonical}).

To separate the constraints into first and second-classes, we
first calculate the commutation relations of (\ref{eqq}) to yield
the nonvanishing Poisson brackets \bea
\{C^{(1)}_{ab}(x),C^{(4)}_{cd}(y)\}&=
&\frac{1}{g^2}(r_c-r_d)\delta_{ad} \delta_{bc}\delta(x-y),\label{c14}\\
\{C^{(2)}_{ab}(x),C^{(3)}_{cd}(y)\}&=&(r_c-r_d)\delta_{ad}\delta_{bc}\delta(x-y),\label{c23}\\
\{C^{(2)}_{ab}(x),C^{(4)}_{cd}(y)\}&=&(r_a+r_b)\delta_{ad}\delta_{bc}\delta(x-y),\label{c24}\\
\{C^{(3)}_{ab}(x),C^{(3)}_{cd}(y)\}&=&\left(\delta_{bc}C^{(3)}_{ad}
-\delta_{ad}C^{(3)}_{cb}\right)\delta(x-y), \label{c33}\\
\{C^{(3)}_{ab}(x),C^{(4)}_{cd}(y)\}&=&
\frac{1}{g^2}([A_0, R]_{ad}\delta_{bc}-[A_0, R]_{bc}
\delta_{ad})\delta(x-y),\label{c34}\\
\{C^{(4)}_{ab}(x),C^{(4)}_{cd}(y)\}&=
&\kappa(F_{12cb}\delta_{ad}-F_{12ad}\delta_{bc})\delta(x-y).
\label{c44} \eea
Note that (\ref{c33}) satisfies $U(2)$ Gauss law algebra.
Nevertheless, $C^{(3)}_{12}$ and $C^{(3)}_{21}$ become second
class constraints for $r\neq 1$, because in this case the right
hand sides of (\ref{c23}) and (\ref{c34}) are nonvanishing for $c
\neq d$.

Before proceeding to the calculation of the Dirac brackets we
briefly review in the next section the structure of the
constrained phase space in a geometric language. This section is
included mainly to fix our notations, conventions and terminology.

\section{Geometry of Constrained Phase Space}
\setcounter{equation}{0}
\renewcommand{\theequation}{\arabic{section}.\arabic{equation}}

A phase space can be described by a manifold $\Gamma$ with a
non-degenerate closed $2$-form, $\Omega_{AB}$. The capital Roman
letters $(A,B\cdots)$ are used to represent collectively the
indices of the phase space coordinates. In our case
$x^{A}=(\Pi^\alpha_a, \psi^\alpha_a, A_{i ab}, A_{0 ab}, P_{0ab},
\lambda_{ab}, \Pi^{\lambda}_{ab})$. The Poisson bracket structure
on $\Gamma$ is defined as follows. For any given two functions
$F$, $G$ \beq \{F,G\}=\Omega^{AB}\partial_{A} F\partial_{B}
G,\label{PB} \eeq where $\Omega^{AB}$ denotes the inverse of
$\Omega_{AB}$.

If a theory is constrained by the constraints, $C^{\bar\mu}=0$,
the space of physical interests will be the submanifold
$\bar\Gamma$ consisting of all points of $\Gamma$ satisfying the
constraints. This constrained subspace inherits a closed $2$-form,
$\bar\Omega_{AB}$, from $\Omega_{AB}$ by restriction, i.e., for
any two vector fields ${\bar X}^{A}$, ${\bar Y}^{B}$ tangent to
$\bar\Gamma$ we define $\bar\Omega_{AB}$ by \beq
\bar\Omega_{AB}{\bar X}^{A} {\bar Y}^{B}\equiv \Omega_{AB}{\bar
X}^{A} {\bar Y}^{B}. \eeq Let us divide the discussion in two
cases.

1. $\bar\Omega_{AB}$ is non-degenerate.

In this case, ($\bar\Gamma$, $\bar\Omega_{AB}$) is the reduced
phase space and the reduced bracket structure can be defined as
before, using the inverse of $\bar\Omega_{AB}$. For any two
functions $\bar F$, $\bar G$ of $\bar\Gamma$ we define \beq \{\bar
F,\bar G\}_{D}={\bar\Omega}^{AB}\partial_{A} {\bar
F}\partial_{B}{\bar G}. \eeq The condition for non-degeneracy of
$\bar\Omega_{AB}$ can be stated as \beq
\rm{det}\{C^{\bar\mu},C^{\bar\nu}\}\ne 0.\eeq This condition, in
turn, is equivalent to the fact that none of the vectors
$\Omega^{AB}\partial_{B} C^{\bar\mu}$ is tangent to $\bar\Gamma$.
In this case, the constraints $C^{\bar \mu}=0$ are said to form a
second class and the resulting bracket structure on $\bar\Gamma$
is called the Dirac bracket to distinguish it from the original
Poisson bracket, (\ref{PB}).

It is well known that $\bar\Omega^{AB}$, when regarded as a tensor
field of $\Gamma$, both of whose indices are tangent to the
submanifold $\bar\Gamma$, is related to $\Omega^{AB}$ as follows.
\beq {\bar\Omega}^{AB}= \Omega^{AB}+
\Theta^{-1}_{\bar\mu\bar\nu}\Omega^{AM}\partial_{M} C^{\bar\mu}
\Omega^{BN}\partial_{N} C^{\bar\nu}, \label{inverse} \eeq where
$\Theta^{\bar\mu\bar\nu}\equiv\{C^{\bar\mu},C^{\bar\nu}\}.$ In
terms of the Poisson bracket, the Dirac bracket can be written as
\beq \{F,G\}_{D}=
\{F,G\}-\{F,C^{\bar\mu}\}\Theta^{-1}_{\bar\mu\bar\nu}\{C^{\bar\nu},G\}.
\label{diracbra}\eeq

2. $\bar\Omega_{AB}$ is degenerate.

The situation in this case is slightly more complicated because
the inverse does not exist. Therefore, we cannot define the
bracket structure on all of the functions of $\bar\Gamma$.
However, $\bar\Omega_{AB}$ defines for us a non-degenerate closed
$2$-form on the quotient manifold of $\bar\Gamma$ where any two
points of $\bar\Gamma$ are identified if they are related by a
curve which lies along the degeneracy directions everywhere. In
fact, $\bar\Omega_{AB}$ is the pull-back to $\bar\Gamma$ of a
non-degenerate closed $2$-form on the quotient space under the
quotient map. We will interpret $\bar\Omega_{AB}$ in both ways,
either as a degenerate $2$-form on $\bar\Gamma$ or as a
non-degenerate $2$-form on the quotient manifold. In this case,
the quotient space together with a non-degenerate closed $2$-form,
$\bar\Omega_{AB}$, is the fully reduced phase space and one can
define the bracket structure. Physically, the degeneracy
directions represent gauge directions and the quotient space is
the space of gauge orbits. Since gauge invariant functions can be
identified with the functions on the quotient manifold, the fact
that we have a well defined bracket structure on the quotient
space means that the bracket structure can be well defined only on
gauge invariant functions on $\bar\Gamma$.

Degeneracies are in fact associated with the existence of the
so-called first class constraints. Let $k^{A}$ be an arbitrary
vector field on $\bar\Gamma$ which points in some degeneracy
direction. Then, for all vector fields, $t^{B}$, tangent to
$\bar\Gamma$, \beq 0=\bar\Omega_{AB}k^{A} t^{B}= \Omega_{AB}k^{A}
t^{B}, \eeq which implies that \beq
\Omega_{AB}k^{A}=\lambda_{\bar\mu}
\partial_{B} C^{\bar\mu}=
\partial_{B} (\lambda_{\bar\mu} C^{\bar\mu})\eeq
for some non-trivial $\lambda_{\bar\mu}$. Such a linear
combination of the constraints, $\lambda_{\bar\mu} C^{\bar\mu}$,
is called a first class constraint and its Poisson bracket with
all other constraints vanishes, i.e., \beq \{\lambda_{\bar\mu}
C^{\bar\mu}, C^{\bar\nu}\}=0. \eeq Conversely, when
$\Theta^{\bar\mu\bar\nu}\equiv \{C^{\bar\mu},C^{\bar\nu}\}$ is
degenerate, there exists a non-trivial $\lambda_{\bar\mu}$ such
that $\lambda_{\bar\mu} \Theta^{\bar\mu\bar\nu}=0$ and it can be
shown that $\lambda_{\bar\mu} C^{\bar\mu}$ generates a degeneracy
of $\bar\Omega_{AB}$. That is,
$k^{A}=\Omega^{AB}\partial_{B}(\lambda_{\bar\mu} C^{\bar\mu})$ is
tangent to $\bar\Gamma$ and $\bar\Omega_{AB}k^{A} t^{B}=0$ for all
$t^{B}$ tangent to $\bar\Gamma$. Other linear combinations of the
constraints independent with all first class constraints belong to
the second class. Therefore, in the degenerate case one can
decompose the constraints into two classes,
$(C^{\bar\mu})=(C^{\bar a}, C^{\bar i})$, where $C^{\bar a}$
denotes the first class constraints and $C^{\bar i}$ the second
class and they satisfy \beq \{C^{\bar a}, C^{\bar\mu}\}=0,~
\rm{det}\{C^{\bar i},C^{\bar j}\}\ne 0. \label{secondc}\eeq

Unlike $\bar\Omega_{AB}$, which can be regarded either as a
non-degenerate $2$-form on the quotient manifold or as a
degenerate $2$-form on $\bar\Gamma$, $\bar\Omega^{AB}$ has a well
defined meaning only as a tensor field on the quotient space. In
order to compare it with $\Omega^{AB}$ we choose a gauge slice.
Then, using this one to one map between the space of gauge orbits
and the gauge slice one obtains the corresponding non-degenerate
closed $2$-form and its inverse on the gauge slice. Note that the
$2$-form on the gauge slice obtained this way is just the induced
$2$-form from $\Omega_{AB}$ by restriction to the gauge slice.
Therefore, one can obtain the relations between $\bar\Omega^{AB}$
and $\Omega^{AB}$ by treating the gauge slicing conditions as
additional constraints. When these are included all constraints
form a second class as one can see from the fact that the induced
$2$-form on the gauge slice is non-degenerate. Let $G^{\bar a}=0$
represent a choice of gauge slice. For this to be a good choice of
gauge slicing $W^{\bar a \bar b}\equiv\{G^{\bar a},C^{\bar b}\}$
should be invertible. Then, from Eq. (\ref{diracbra}) one obtains,
after a straightforward calculation,
\begin{eqnarray}
\{F,G\}^{\prime}_D&\equiv&{\bar\Omega}^{AB}\partial_{A}F\partial_{B}G\nonumber\\
&=& \{F,G\}\nonumber\\
&+& W^{-1}_{\bar a\bar m }W^{-1}_{\bar b\bar n }\left(\{G^{\bar
m},G^{\bar n}\}-\{G^{\bar m}, C^{\bar i}\}\{G^{\bar n}, C^{\bar
j}\}\Theta^{-1}_{\bar i \bar j}\right)\{C^{\bar a},F\}\{C^{\bar
b},G\}\nonumber\\
&+&W^{-1}_{\bar a \bar b}\{C^{\bar a},F\}\{G^{\bar b},G\}
-W^{-1}_{\bar a \bar b}\{G^{\bar b},F\}\{C^{\bar a},G\}\nonumber\\
&+&W^{-1}_{\bar a \bar b}\{G^{\bar b}, C^{\bar
i}\}\Theta^{-1}_{\bar i \bar j}\{C^{\bar j},F\}\{C^{\bar a},G\}
-W^{-1}_{\bar a \bar b}\{G^{\bar b}, C^{\bar i}\}\Theta^{-1}_{\bar
i \bar j}\{C^{\bar a},F\}
\{C^{\bar j},G\}\nonumber\\
&-& \Theta^{-1}_{\bar i \bar j}\{C^{\bar i},F\}\{C^{\bar j},G\},
\label{dbra}
\end{eqnarray}
where $\Theta^{\bar i\bar j}=\{C^{\bar i},C^{\bar j}\}$. When the
functions $F$, $G$ are gauge invariant the above equation reduces
to the usual Dirac bracket constructed using the second class
constraints only.

From geometric point of view what happens in our model can be
explained as follows. The vector fields which are
(Poisson-)generated by the non-diagonal part of $U(2)$ constraints
point in fixed directions in $\Gamma$. When $r\ne 1$, they are not
tangent to $\bar\Gamma$. As the parameter, $r$, approaches one,
the constraints change gradually and $\bar\Gamma$ becomes tangent
to those vector fields at $r=1$. Initially second class
constraints become first class, the gauge symmetry being enlarged
from $U(1)\times U(1)$ to $U(2)$.

\section{Dirac Brackets}
\setcounter{equation}{0}
\renewcommand{\theequation}{\arabic{section}.\arabic{equation}}

In this section, we explicitly construct the Dirac brackets
(\ref{diracbra}) of our model.
It turns out that transition from $r\neq 1$ to $r=1$ is singular
and we have to carry out the cases of $r=1$ and $r\neq 1$
separately. The reason is that in the Dirac method we have to
compute the inverse of the Dirac matrix $\Theta^{\bar i\bar j}$ of
(\ref{inverse}) which is constructed with second class constraints
only. This inverse matrix becomes singular in
the limit of $r \rightarrow 1$, because part of the constraints
change from second class into first class in the limit, and
determinant of the Dirac matrix becomes zero.

\subsection{$r=1$ case}

For the case of $r=1$, we have $R_{ab}=\delta_{ab}$, and it is
easy to infer from the constraints algebra
(\ref{c14})-(\ref{c44}), only $C^{(2)}_{ab}$ and $C^{(4)}_{ab}$
are second class constraints. All of $C^{(3)}_{ab}$'s are the
first class constraints whose Gauss law satisfies the $U(2)$
algebra (\ref{c33}).
$C^{(0)}_{ab}$ and $C^{(1)}_{ab}$ completely decouple from the
theory and can be put equal to zero.

One can thus obtain the following Poisson bracket relations
$\Theta^{\bar i\bar j}=\{C^{\bar i},C^{\bar j}\}$ among the
second-class constraints $C^{\bar
i}=(C^{(2)}_{11},C^{(2)}_{12},C^{(2)}_{21},C^{(2)}_{22},
C^{(4)}_{11},C^{(4)}_{12},C^{(4)}_{21},C^{(4)}_{22})$ $({\bar i}
=1,2,...,8)$, \beq \Theta= \left[
\begin{array}{cc}
O &M\\
-M^{T} &N\\
\end{array}
\right] \label{thetaab} \eeq where
\beq M=\left[
\begin{array}{cccc}
2g_{11} &0  &0  &0\\
0  &0 &2g_{11} &0\\
0  &2g_{11} &0 &0\\
0  &0  &0 &2g_{11}\\
\end{array}
\right],~~~ \ N=\left[
\begin{array}{cccc}
0 &-f_{12} &f_{21} &0\\
f_{12} &0  &-\delta f  &-f_{12}\\
-f_{21} &\delta f  &0  &f_{21}\\
0 &f_{12}  &-f_{21} &0\\
\end{array}
\right]. \label{matrixmn0} \eeq Here we have defined,
$g_{11}\equiv|\psi_{1}|^{2}=r$,
$g_{22}\equiv|\psi_{2}|^{2}=r^{-1}$, $f_{ab}=\kappa F_{12ab}$ and
$\delta f=f_{11}-f_{22}$. For $r=1$ we have $g_{11}=g_{22}=1$

The inverse matrix of $\Theta$ is given by
\begin{equation}
\Theta^{-1}=\left[
\begin{array}{cc}
M^{T-1}N M^{-1} &-M^{T-1}\\
M^{-1}          &O\\
\end{array}
\right],
\label{thetainverse0}
\end{equation}
with \beq M^{-1} = \left[
\begin{array}{cccc}
\frac{1}{2g_{11}} &0 &0 &0\\
0 &0 &\frac{1}{2g_{11}} &0\\
0 &\frac{1}{2g_{11}} &0 &0\\
0 &0 &0 &\frac{1}{2g_{11}}\\
\end{array}
\right],~~~ M^{T-1}NM^{-1}=\left[
\begin{array}{cccc}
0 &\frac{f_{21}}{4g_{11}^{2}} &-\frac{f_{12}}{4g_{11}^{2}} &0\\
-\frac{f_{21}}{4g_{11}^{2}} &0 &\frac{\delta f}{4g_{11}^{2}}
&\frac{f_{21}}{4g_{11}^{2}}\\
\frac{f_{12}}{4g_{11}^{2}} &-\frac{\delta f}{4g_{11}^{2}} &0
&-\frac{f_{12}}{4g_{11}^{2}}\\
0 &-\frac{f_{21}}{4g_{11}^{2}} &\frac{f_{12}}{4g_{11}^{2}} &0\\
\end{array}
\right]. \label{mninverse} \eeq
The Dirac brackets (\ref{diracbra}) are then given by \bea
\{\psi_a^\alpha (x),\Pi_b^\beta(y)\}_{D}&=&
\left(\delta_{ab}\delta^{\alpha\beta}
-\frac{\psi_{c}^{\alpha}\psi_{c}^{\beta\dagger}}{2g_{11}}\delta_{a1}\delta_{b1}
+(1\leftrightarrow 2)\right)\delta(x-y),\nonumber\\
\{\psi_a^{\alpha\dagger} (x),\Pi_b^{\beta}(y)\}_{D}&=&\left(
-\frac{\psi_{1}^{\alpha\dagger}\psi_{1}^{\beta\dagger}}{2g_{11}}\delta_{a1}\delta_{b1}
-\frac{\psi_{2}^{\alpha\dagger}\psi_{1}^{\beta\dagger}}{2g_{11}}\delta_{a1}\delta_{b2}
+(1\leftrightarrow 2)\right)\delta(x-y),\nonumber\\
\{\Pi_a^{\alpha} (x),\Pi_b^{\beta}(y)\}_{D}
&=&\left[\left(\frac{\Pi_{1}^{\alpha}\psi_{1}^{\beta\dagger}-\psi_{1}^{\alpha\dagger}\Pi_{1}^{\beta}}{2g_{11}}
+\frac{\psi_{2}^{\alpha\dagger}\psi_{1}^{\beta\dagger}
-\psi_{1}^{\alpha\dagger}\psi_{2}^{\beta\dagger}}{2g_{11}}\frac{f_{12}}{2g_{11}}
\right) \delta_{a1}\delta_{b1}
+\left(\frac{\Pi_{2}^{\alpha}\psi_{1}^{\beta\dagger}-\psi_{2}^{\alpha\dagger}\Pi_{1}^{\beta}}{2g_{11}}
\right.\right.\nonumber\\
& &\left.\left.
+\frac{\psi_{1}^{\alpha\dagger}\psi_{1}^{\beta\dagger}}{2g_{11}}\frac{f_{21}}{2g_{11}}
-\frac{\psi_{2}^{\alpha\dagger}\psi_{2}^{\beta\dagger}}{2g_{11}}\frac{f_{12}}{2g_{11}}
-\frac{\psi_{2}^{\alpha\dagger}\psi_{1}^{\beta\dagger}}{2g_{11}}\frac{\delta
f}{2}\right)\delta_{a1}\delta_{b2}+(1\leftrightarrow 2)
\right]\delta(x-y),\nonumber\\
\{\Pi_a^{\alpha} (x),\Pi_b^{\beta\dagger}(y)\}_{D}
&=&\left[\left(\frac{\Pi_{c}^{\alpha}\psi_{c}^{\beta}-\psi_{c}^{\alpha\dagger}
\Pi_{c}^{\beta\dagger}}{2g_{11}}
+\frac{\psi_{2}^{\alpha\dagger}\psi_{1}^{\beta\dagger}}{2g_{11}}\frac{f_{12}}{2g_{11}}
+\frac{\psi_{1}^{\alpha\dagger}\psi_{2}^{\beta\dagger}}{2g_{11}}\frac{f_{21}}{2g_{11}}
-\frac{\psi_{2}^{\alpha\dagger}\psi_{2}^{\beta}}{2g_{11}}\frac{\delta
f}{2} \right)\delta_{a1}\delta_{b1}
\right.\nonumber\\
& &\left.
-\frac{\psi_{c}^{\alpha\dagger}\psi_{c}^{\beta}}{2g_{11}}\frac{f_{12}}{2g_{11}}
\delta_{a1}\delta_{b2}
+(1\leftrightarrow 2)\right]\delta(x-y),\nonumber\\
\{\lambda_{ab}(x),\Pi_{cd}^{\lambda}(y)\}_{D}
&=&\delta_{ac}\delta_{bd}\delta(x-y),\nonumber\\
\{A_{0ab}(x),P_{0cd}(y)\}_{D}&=&\delta_{ac}\delta_{bd}\delta(x-y),\nonumber\\
\{A_{iab}(x),A_{jcd}(y)\}_{D}&=&-\frac{1}{\kappa}\epsilon_{ij}
\delta_{ad}\delta_{bc}\delta(x-y). \label{diracbracket0} \eea

\subsection{$r\neq 1$ case}
In this case, we first note that two of the constraints
$C^{(3)}_{12}$ and $C^{(3)}_{21}$ which were first-class in the
case of $r=1$ become second-class, because the gauge symmetry is
reduced to $U(1)\times U(1)$. This is evident from (\ref{c14}),
whose right hand side is nonvanishing  for $r_c \neq r_d$.
Therefore, we have all together twelve second class constraints
($C^{(1)}_{12}, C^{(1)}_{21}, C^{(2)}_{11}, C^{(2)}_{12},
C^{(2)}_{21}, C^{(2)}_{22},C^{(3)}_{12},
C^{(3)}_{21},C^{(4)}_{11},
C^{(4)}_{12},C^{(4)}_{21},C^{(4)}_{22}).$ One could proceed to the
computation of the Dirac bracket with these twelve constraints,
which is quite involved. However, it greatly simplifies the
computation if one observes that the constraints $C^{(4)}_{12}$
and $C^{(4)}_{21}$ can be eliminated from the list by solving them
explicitly with the variables $A_{0ab}~(a\neq b) $given by \beq
A_{0ab}=\frac{g^2}{r_b-r_a}(\Pi_a \psi_b+\psi^\dagger_a
\Pi^\dagger_b) ~(a\neq b). \eeq Then, from
(\ref{c14})-(\ref{c44}),  $C^{(1)}_{12}$ and $C^{(1)}_{21}$
commutes with the rest of the costraints, and the number of the
second-class constraints become eight; $C^{\bar
i}=(C^{(2)}_{11},C^{(2)}_{12},C^{(2)}_{21},C^{(2)}_{22},C^{(3)}_{12},
C^{(3)}_{21}$,$C^{(4)}_{11},C^{(4)}_{22})$, $({\bar i}=1,...,8)$.

We now find a $8\times 8$ matrix $\Theta^{\bar i\bar j}=\{C^{\bar
i},C^{\bar j}\}$ of the form
\begin{equation}
\Theta=\left[
\begin{array}{cc}
O &M\\
-M^{T}  &0\\
\end{array}
\right], \label{theta2}
\end{equation}
where $M$ is given by \beq M=\left[ \matrix{0 &0 &2g_{11} &0 \cr 0
&\delta g &0 &0 \cr -\delta g &0 &0 &0\cr 0 &0 &0 &2g_{22} \cr}
\right], \label{matrixmn2} \eeq with $\delta g=g_{11}-g_{22}$. The
inverse matrix of $\Theta$ is given by
\begin{equation}
\Theta^{-1}=\left[
\begin{array}{cc}
O &-(M^{-1})^{T}\\
M^{-1}  &0\\
\end{array}
\right],
\label{thetainverse2}
\end{equation}
with \beq M^{-1} = \left[
\begin{array}{cccc}
0 &0 &-\frac{1}{\delta g} &0\\
0 &\frac{1}{\delta g} &0 &0\\
\frac{1}{2g_{11}} &0 &0 &0\\
0 &0 &0 &\frac{1}{2g_{22}}\\
\end{array}
\right]. \label{mninverse2} \eeq
The Dirac bracket is then given by \bea
\{\psi_{a}^{\alpha}(x),\Pi_{b}^{\beta}(y)\}_{D}&=&
\left[\delta_{ab}\delta^{\alpha\beta}
+\left(-\frac{\psi_{1}^{\alpha}\psi_{1}^{\beta\dagger}}{2g_{11}}
+\frac{\psi_{2}^{\alpha}\psi_{2}^{\beta\dagger}}{\delta g}
\right)\delta_{a1}\delta_{b1}
+(1\leftrightarrow 2)\right]\delta(x-y),\nonumber\\
\{\psi_a^{\alpha\dagger}(x),\Pi_{b}^{\beta}(y)\}_{D}&=&\left[
-\frac{\psi_{1}^{\alpha\dagger}\psi_{1}^{\beta\dagger}}{2g_{11}}
\delta_{a1}\delta_{b1}
+\frac{\psi_{2}^{\alpha\dagger}\psi_{1}^{\beta\dagger}}{\delta g}
\delta_{a1}\delta_{b2}
+(1\leftrightarrow 2)\right]\delta(x-y),\nonumber\\
\{\Pi_a^{\alpha}
(x),\Pi_b^{\beta}(y)\}_{D}&=&\left[\frac{\Pi_{1}^{\alpha}\psi_{1}^{\beta\dagger}
-\psi_{1}^{\alpha\dagger}\Pi_{1}^{\beta}}{2g_{11}}
\delta_{a1}\delta_{b1}
+\frac{\Pi_{2}^{\alpha}\psi_{1}^{\beta\dagger}+\psi_{2}^{\alpha\dagger}
\Pi_{1}^{\beta}}{\delta g}
+(1\leftrightarrow 2)\right]\delta(x-y),\nonumber\\
\{\Pi_a^{\alpha} (x),\Pi_b^{\beta\dagger}(y)\}_{D}&=&\left[\left(
\frac{\Pi_{1}^{\alpha}\psi_{1}^{\beta}-\psi_{1}^{\alpha\dagger}\Pi_{1}^{\beta\dagger}}{2g_{11}}
+\frac{\Pi_{2}^{\alpha}\psi_{2}^{\beta}-\psi_{2}^{\alpha\dagger}\Pi_{2}^{\beta\dagger}}
{\delta g}\right) \delta_{a1}\delta_{b1}
+(1\leftrightarrow 2)\right]\delta(x-y),\nonumber\\
\{A_{iab}(x),A_{jcd}(y)\}_{D}&=&-\frac{1}{\kappa}\epsilon_{ij}\delta_{ad}\delta_{bc}
\delta(x-y). \label{diracbracket2} \eea We note that not only the
structure of constraints is different from $r=1$ case, but also
$r\rightarrow 1$ is not defined in the above algebra
(\ref{diracbracket2}).

\section{Conclusion}
\setcounter{equation}{0}
\renewcommand{\theequation}{\arabic{section}.\arabic{equation}}
We performed canonical analysis of the gauge symmetry enhancement
in the enlarged $CP(N)$ model coupled with $U(2)$ Chern-Simons
term. We discussed  the transition between $r=1$ and $r\neq 1$
cases in terms of the degeneracy of the constrained phase space
geometry. We found that the conventional Dirac method does not
allow a smooth extrapolation of the symmetry enhanced and broken
phases. This was essentially due to the fact that Dirac procedure
requires an inverse of the Dirac matrix which is constructed with
second class constraints only, and becomes singular when some of
the second class constraints become first class. Physically,
second order phase transition occurring as the symmetry breaking
parameter $r$ approaches the critical value 1 could be responsible
for the non-smooth transition.

We conclude with a couple of remarks. We have computed the Dirac
bracket of (\ref{diracbra}) without gauge fixing and thus are
considering only gauge invariant functions which commutes with
the first class constraints. Instead one could try to fix the
gauge first thereby rendering all the constraints second class,
and then proceed to the Dirac bracket (\ref{dbra}). This would
involve technically more difficult steps; for example, in the
case of $r=1$, we need four gauge fixing conditions corresponding
to the $U(2)$ gauge symmetry, which could be chosen as Lorentz
gauge. Then the matrix would become $16\times 16$. For the gauge
conditions corresponding to $U(1)\times U(1)$ in the case of
$r\neq 1$, we have to evaluate the inverse of  $12\times 12.$
Finally, it would  be interesting to perform other quantization
methods of our model.  For example, in the BRST-BFV method
\cite{bfv} which avoids the second class constraints from the
beginning by enlarging the phase space, the issue of the
connection between  $r=1$ and $r\neq1$  values could be
reexamined.

\acknowledgments
STH would like to acknowledge financial support
in part from the Korea Science and Engineering Foundation Grant
R01-2000-00015. PO was supported by Korea Research Foundation grant
(KRF-2002-042-C00010).

\end{document}